\documentclass[12pt]{iopart}
\usepackage{graphics}
\usepackage{cite}
\usepackage{here}

\newcommand{\bal}{\mbox{\boldmath $\alpha$}}
\newcommand{\brv}{\mbox{\boldmath $r$}}
\newcommand{\bpv}{\mbox{\boldmath $p$}}
\newcommand{\bjv}{\mbox{\boldmath $j$}}
\newcommand{\bLv}{\mbox{\boldmath $L$}}
\newcommand{\bsiv}{\mbox{\boldmath $\sigma$}}

\begin{document}

\title[Spin dynamics of WP in strong Coulomb field]
{Spin dynamics of wave packets evolving with the Dirac Hamiltonian in 
atoms with high Z}
\author{R Arvieu\dag ~~P Rozmej\ddag  
~~and ~M Turek\ddag\footnote[4]{E-mail: rozmej@tytan.umcs.lublin.pl,
mturek@curie.umcs.lublin.pl and arvieu@in2p3.fr}} 

\address{\ddag Institut des Sciences Nucl\'eaires,  F-38026 Grenoble,  
France}
\address{\dag Instytut Fizyki,  Uniwersytet MCS,  20-031 Lublin,  
Poland}

\begin{abstract}
The motion of circular WP for one electron in central Coulomb
field with high $Z$ is calculated. The WP is defined in terms of solutions 
of the Dirac equation in order to take into account all possible relevant
effects in particular the spin-orbit potential. A time scale is defined
within which spin dynamics must be taken into account mainly in the atoms
with high $Z$. Within this time scale there exists a mechanism of collapses
and revivals of the spin already shown by the authors for harmonic
oscillator potential and called the {\em spin orbit pendulum}. However this
effect has not the exact periodicity of the simpler model, but the WP's spatial
motion is nevertheless quite similar.
\end{abstract}
\pacs{03.65.Sq} 

\vspace{2mm}
\noindent Submitted to {\em Physical Review A}   \hfill\today

\section{Introduction}\label{intro}
In the last decade large efforts
\cite{parker,averbukh,nauenberg,dacic,peres,bluhmsr,BKP} 
have been
devoted to the detailed understanding of the quantum dynamics of WP in
simple systems like H atoms or hydrogenoid atoms.
Theoretical investigations resulted in very good understanding
of such subtle interference effects like collapse, revivals,
fractional revivals and super-revivals of wavepackets created
in atomic and molecular systems. Many signatures of these phenomena
have been observed in numerous sophisticated experimental studies.
Indeed with contemporary lasers
and tunable electromagnetic fields one can engineer many different
desired initial states and analyze their time evolution \cite{alber}.
   
     Our aim in this paper is to perform a new step in the analysis of the
WP dynamics for hydrogen atom and for ions with one electron by
considering spin motion. Until now this effect was not considered for
isolated atoms since it requires very long time scales to manifest itself
due to the weakness of the spin--orbit coupling. In order to decrease as
much as possible this time scale we will consider WP within ions with very
large $Z$. We intended to make a calculation which include all possible
effects for these elements and therefore we have considered WP which
evolve under the relativistic Dirac Hamiltonian i.e. which are 
four component spinors.

    Spin effects and relativistic effects have been studied many times but
mostly for atoms submitted to intense laser fields. A recent review has
been given in \cite{protopapas}.
Some of the most relevant research works are found in
\cite{latinne,rathe,szymanowski,husx}.
It is generally agreed in them that spin and relativistic effects
are small but that their magnitude increase with the intensity of the
field. For fields of the order of $10^{16}$ to $10^{17}$ W/cm$^2$ 
they should be taken into account. 
In a most recent paper \cite{husx} the interest has been
focussed on the spin--orbit coupling, essentially for Al$^{+12}$ and Ga$^{+12}$
ions. The dynamics of the spin degree of freedom was analyzed there during
the time of interaction of the laser with the ion. The authors did not use
the Dirac equation but the Schr\"odinger equation corrected up to order
$1/c^3$ and therefore the spin--orbit potential could be clearly identified
by its mere suppression. The most significant signature occur in the
radiation spectrum. However little polarization was produced for the ions
which were considered.  

    Spin dynamics was previously studied by us in harmonic oscillator
potentials in non-relativistic dynamics. In ref.\ \cite{arv94,arv95,roz96}  
we have found a new
effect that we have called the {\em spin--orbit pendulum} for a spin--orbit
potential with a constant radial part. This phenomenon is simply the
periodic collapse and revival of the average spin and an exchange of
angular momentum between spin and the orbital angular momentum. 
The analysis is quite parallel to that of the Jaynes-Cummings model 
\cite{jaynes}. In order to extend this
situation to a relativistic dynamics it was quite natural to study the
Dirac oscillator (DO). There, the strong entanglement that gives rise to
spin oscillation is still present \cite{roz99}. 
Because of nonlinearities of the relativistic energies the dynamics is not 
periodic. We have also studied relativistic effects like zitterbewegung. 
Our studies have been also extended to WP motion corresponding to linear
trajectories in non-relativistic approach (HO+LS Hamiltonian) \cite{arv97}
as well as in the relativistic Dirac oscillator \cite{tur00}.
 
     Our previous studies stimulated the present work for hydrogen or for
heavy ions. It is expected that many features of the DO should take place
in this much more realistic situation.
   
Our paper is organized as follows. In order to make it self-contained
we in Sec.~\ref{dirac} present basic formula for solutions
of Dirac equation with Coulomb potential. In Sec.~\ref{construction} 
we show the construction of the circular WP and in Sec.\ 
\ref{evolution} we discuss its time evolution, focusing on
the autocorrelation function, spin expectation values 
and spatial motion. Particular attention is paid to 
relativistic effects and different time scales. 
The concluding remarks are contained in Sec.~\ref{conclusions}.

\section{Dirac equation with central potential}\label{dirac}

Solution of Dirac equation for a central field, in particular
for Coulomb potential is one of standard, textbook problems
in quantum mechanics \cite{rose,greiner}. Therefore we will
recall here only few formula that will be relevant for 
our further considerations.

Dirac equation for a spherical potential $V(r)$ is
\begin{equation}\label{direq}
H_D\Psi(\brv) =(c\hat{\bal}\cdot\bpv + \hat{\beta} mc^2 + V(r))
\Psi(\brv)  = E \Psi(\brv)  \;.
\end{equation}
Its eigenfunctions can be found in a representation that
diagonalizes simultaneously with $H_D$ three other operators,
$\bjv^2, j_z$ and $K=\hbar +\bLv\cdot\bsiv$.  
Then we may write for $\Psi(\brv)$ (using phase convention as in
\cite{greiner})
\begin{equation}\label{psi}
\Psi = \Psi^m_\kappa = \rmi\left( 
\begin{array}{c} g(r)\chi^m_\kappa \\
\rmi f(r)\chi^m_{-\kappa} \end{array} \right)
= \rmi\left(\begin{array}{c} g(r)\Omega_{ljm} \\
\rmi f(r)\Omega_{l'jm} \end{array} \right)
\end{equation}
with 
\begin{equation}\label{lprim}
l' = 2j-l = \left \{ 
\begin{array}{c} 2(l+\case{1}{2})-l=l+1 \hspace{5ex} \mbox{for}
 \hspace{2ex} j= l+\case{1}{2}=j_+  \\
 2(l-\case{1}{2})-l=l-1 \hspace{5ex} \mbox{for}
 \hspace{2ex} j= l-\case{1}{2}=j_- \end{array} \right. \:.
\end{equation}
The explicit forms of the spherical spinors $\Omega_{ljm}$
are:
\numparts
\begin{equation}\label{Omp}
 \Omega_{l,j=l+\case{1}{2},m} = \Omega_{l,j_+,m} =
 \left( \begin{array}{c}
 \sqrt{\frac{j+m}{2j}}\,Y_{l,m-\case{1}{2}} \\
 \sqrt{\frac{j-m}{2j}}\,Y_{l,m+\case{1}{2}} \end{array} \right) 
\end{equation}
and 
\begin{equation}\label{Omm}
 \Omega_{l, j=l-\case{1}{2},m} = \Omega_{l,j_-,m} =
 \left( \begin{array}{c}
 -\sqrt{\frac{j-m+1}{2j+2}}\,Y_{l,m-\case{1}{2}} \\\hspace{1.6ex}
 \sqrt{\frac{j+m+1}{2j+2}}\,Y_{l,m+\case{1}{2}} \end{array} \right) \;.
\end{equation}
\endnumparts
Radial wave functions are given in terms of confluent hypergeometric
functions $F(a,c,x)$
\begin{equation}\label{hgf}
 F(a,c,x) = 1 + \frac{a}{c}x + \frac{a(a+1)}{c(c+1)}\frac{x^2}{2!} 
 + \ldots \;\;.
\end{equation}
Explicit formula are \cite{rose}:
\numparts
\begin{eqnarray}\label{fr}
 f(r) &=& -\frac{\sqrt{2}\,\lambda^{5/2}}{{\mit\Gamma}(2\gamma+1)}
 \left[\frac{{\mit\Gamma}(2\gamma+n'+1)(1-E)}
 {n'!\,\xi(\xi-\lambda/\kappa)}
 \right]^\frac{1}{2} \,(2\lambda r)^{\gamma-1}\, \rme^{-\lambda r} \\
 &\times &[n'F(-n'+1,2\gamma+1,2\lambda r) -(\kappa-\xi/\lambda)
  F(-n',2\gamma+1,2\lambda r)] \nonumber
\end{eqnarray}
and
\begin{eqnarray}\label{gr}
  g(r) &=& \frac{\sqrt{2}\,\lambda^{5/2}}{{\mit\Gamma}(2\gamma+1)}
 \left[\frac{{\mit\Gamma}(2\gamma+n'+1)(1+E)}
 {n'!\,\xi(\xi-\lambda/\kappa)}
 \right]^\frac{1}{2} \,(2\lambda r)^{\gamma-1}\, \rme^{-\lambda r} \\
 &\times & [-n'F(-n'+1,2\gamma+1,2\lambda r) -(\kappa-\xi/\lambda)
  F(-n',2\gamma+1,2\lambda r)] \nonumber  \;,
\end{eqnarray}
\endnumparts
where $\xi = Z\alpha$, $\gamma=\sqrt{\kappa^2-\xi^2}$,
$\kappa^2=(j+1/2)^2$,
$E=[1+(\frac{\xi}{n'+\gamma})^2]^{-1/2}$ and 
$\lambda=\sqrt{1-E^2}$, $\alpha=e^2/\hbar c=1/137.036$ being the {\em
fine structure constant}. 
The nonnegative integer number $n'$ is related to $n$ and $|\kappa|$
by $n=n'+|\kappa|$.
In the following calculations we use atomic units
$m_e=\hbar=c=1$.

\section{Construction of wavepacket}\label{construction}

In our first paper on relativistic WP in Dirac oscillator (DO)
\cite{roz99} we were considering 
a circular spinor without small components written as
\begin{equation}\label{spinor}
 \Psi = \sum_l\, w_l \left( \begin{array}{c} 
  a |lll\rangle \\ b|lll\rangle \\ 0 \\ 0 \end{array} \right)
 =  \sum_l\, w_l \, R_{n=l}(r)|ll\rangle 
 \left( \begin{array}{c} a\\ b \\ 0 \\ 0 \end{array} \right) \;.
\end{equation} 
Here $a$ and $b$ are coefficients which determine the 
initial direction of the spin.
There is no loss of generality to choose them real. $R_{n=l}(r)$
denotes radial wave function of the harmonic oscillator. The large
component for the partial wave $l$ is written as
\begin{equation}\label{spll}
 |ll\rangle \left( \begin{array}{c}a\\ 
 b \end{array}\right)
 = a \Omega_{l,j_+,j_+} + b \left( 
 \frac{\Omega_{l,j_+,j_-} + \sqrt{2l}\,\Omega_{l,j_-,j_-}}
 {\sqrt{2l+1}} \right) \;.
\end{equation}
Weights $w_l$ of the superposition (\ref{spinor}) 
were chosen in a way providing $\Psi$ to be the HO coherent state. 
For DO the bispinors without small components were constructed as 
a superposition of spinors with positive and negative energies
that provide a cancellation of the small components.

For the Dirac equation with Coulomb potential we are not able 
to get rid of small components completely, as we want to build
the WP only from bound states. We can, however, construct
small components in a way similar to the big ones providing
their circular motion. For
this purpose we write the angular parts of the large components as in
eq.~(\ref{spll}), then we complete each of the spinors 
$\Omega_{l,j_+,j_+}, \Omega_{l,j_+,j_-}$ and $\Omega_{l,j_-,j_-}$ 
by their corresponding small components associating each with its
radial part different for the states with $j_+$ and $j_-$ as in 
eq.~(\ref{psi}).
Thus there are four radial parts respectively called 
$g_{j_+}, f_{j_+},g_{j_-}$ and $f_{j_-}$ for a given $l$.
In order to produce circular waves for hydrogenoid atoms the
connection between $n$ and $l$ is $l=n-1$. 

A partial wave at time $t$ can be written with these notations using 
$E_{j_+}$ and $E_{j_-}$ the eigenvalues of the Dirac equation:
\begin{eqnarray}\label{e9}
 |\psi_l(t)\rangle & = & \left[ a \left( \begin{array}{c}
  \rmi\, g_{j_+}\, \Omega_{l,j_+,j_+} \\ -f_{j_+}\,\Omega_{l+1,j_+,j_+}
  \end{array}\right) + \frac{b}{\sqrt{2l+1}} \left( \begin{array}{c}
  \rmi\, g_{j_+}\, \Omega_{l,j_+,j_-} \\ -f_{j_+}\,\Omega_{l+1,j_+,j_-}
  \end{array}\right)\right] \rme^{-\rmi E_+ t} \nonumber \\
  && +  b \sqrt{\frac{2l}{2l+1}} \left( \begin{array}{c}
  \rmi\, g_{j_-}\, \Omega_{l,j_-,j_-} \\ -f_{j_+}\,\Omega_{l-1,j_-,j_-}
  \end{array}\right) \rme^{-\rmi E_- t} \;.
\end{eqnarray}  

 After expansion of the $\Omega$ as in (\ref{Omp},\ref{Omm})
the {\em circular} wavepacket at time $t$
can be written as 
\begin{equation}\label{cirwp}
 \Psi(t) = \left( \begin{array}{c} |c_1(t)\rangle \\
 |c_2(t)\rangle \\|c_3(t)\rangle \\|c_4(t)\rangle  \end{array} \right) \;.
\end{equation} 

For Coulomb potential weights $w_n$
(that can be made real) are chosen in a standard
way  \cite{brown,nauenberg,BKP}, i.e.\ as a Gaussian distribution with 
respect to the mean value $N$:
\begin{equation}\label{gausw}
|w_n|^2 = (2\pi\sigma_G^2)^{-\frac{1}{2}}\,
\rme^{-(n-N)^2/2\sigma_G^2}\;.
\end{equation} 
 
Explicit form of the wavepacket (\ref{cirwp}) is obtained 
in the following form:
\numparts \label{ct14}
\begin{eqnarray}\label{ct1}
\fl 
|c_1(t)\rangle = \rmi \sum_l\,w_l \left \{ g_{j_+}\left[ 
 a |l,l\rangle
  + b \frac{\sqrt{2l}}{2l+1}|l,l-1\rangle \right] \rme^{-\rmi E_{j_+}t}
  \right.  \\ \hspace{10ex} \left.
  -b g_{j_-} \frac{\sqrt{2l}}{2l+1}|l,l-1\rangle \rme^{-\rmi E_{j_-}t}
  \right \} \nonumber
\end{eqnarray}  
\begin{equation}\label{ct2}
\fl 
|c_2(t)\rangle = \rmi \sum_l\,w_l\,b|l,l\rangle \left\{ g_{j_+} 
 \frac{1}{2l+1} \rme^{-\rmi E_{j_+}t}+g_{j_-} \frac{2l}{2l+1} 
 \rme^{-\rmi E_{j_-}t}  \right \}  
\end{equation}  
\begin{eqnarray}\label{ct3}
\fl 
|c_3(t)\rangle = {~} \sum_l\,w_l \left \{ f_{j_+} \left[
  a \frac{1}{\sqrt{2l+3}} |l+1,l\rangle + b \sqrt{\frac{2}
  {(2l+1)(2l+3)}} |l+1,l-1\rangle\right] \rme^{-\rmi E_{j_+}t} \right.
   \nonumber\\
   \hspace{10ex} \left. -b f_{j_-} \sqrt{\frac{2l}{2l+1}}
   |l-1,l-1\rangle \rme^{-\rmi E_{j_-}t} \right \}    
\end{eqnarray} 
\begin{equation}\label{ct4}
\fl 
|c_4(t)\rangle = -\sum_l\,w_l\, f_{j_+} \left \{ 
  a \sqrt{\frac{2l+2}{2l+3}}|l+1,l+1\rangle + b
  \sqrt{\frac{1}{2l+3}}|l+1,l\rangle \right \} \rme^{-\rmi E_{j_+}t} \;.  
\end{equation} 
\endnumparts
 
For $t=0$ spin and orbital angular momentum are only
approximately decoupled, they were exactly decoupled in (\ref{spinor}). 

In numerical calculations we use the full bispinor (\ref{ct1}-\ref{ct4}).
However, particularly for $N>10$, the contributions arising from the
small components ($|c_3\rangle$ and $|c_4\rangle$) are almost negligible,
because the radial $f(r)$ functions are always few orders of magnitude 
smaller than the corresponding $g(r)$ functions.
This is due to the fact that their prefactors differ mainly by the factor
$\sqrt{1-E}$ (few orders of magnitude smaller than 1) for $f(r)$
and $\sqrt{1+E}$ (close to $\sqrt{2}$) for $g(r)$.

\begin{figure}[ht] 
 \resizebox{0.99\textwidth}{!}{\includegraphics{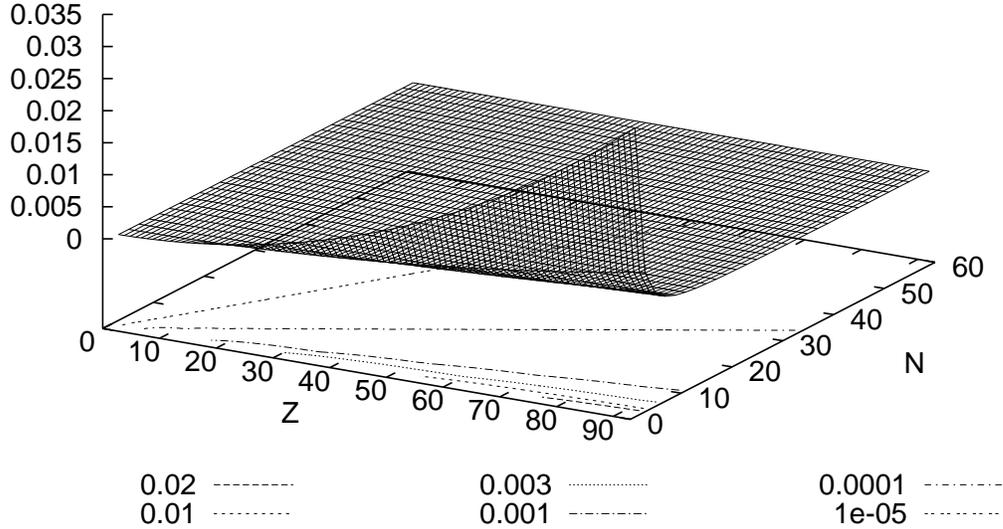} }
\caption{The magnitude of the small components contribution (\ref{small})
as function of $Z$ and $N$.}
\label{smallcontr} 
\end{figure}

Using eqs.~(\ref{ct3}-\ref{ct4}) one can calculate the contributions from
small components of our circular WP exactly. One obtains 
\begin{equation}\label{small}
(\Psi^\dag \Psi)_{small} = \langle c_3(t)|c_3(t) \rangle +
\langle c_4(t)|c_4(t) \rangle
\end{equation}
where 
\numparts \label{c34t}
\begin{equation}\label{c3^2}
\fl \langle c_3(t)|c_3(t)\rangle = \sum_l\, w_l^2 \left \{
 \left[ a^2 \frac{1}{2l+3}+b^2\frac{2}{(2l+1)(2l+3)}\right]
 F_{j_+}  + b^2\frac{2l}{2l+1}F_{j_-}  \right \} \;,
\end{equation} 
\begin{equation}\label{c4^2}
\fl \langle c_4(t)|c_4(t)\rangle = \sum_l\, w_l^2 \left \{
 a^2 \frac{2l+2}{2l+3} + b^2 \frac{1}{2l+3}\right \} F_{j_+} \;. 
\end{equation} 
\endnumparts
In fact, $(\Psi^\dag \Psi)_{small}$ $\langle c_3|c_3\rangle$, and 
$\langle c_4|c_4\rangle$ are constant contributions for our WP.
Terms $F_{j_-}$ and $F_{j_+}$ are radial integrals defined in next
section. Fig.~\ref{smallcontr} displays the magnitude of 
$(\Psi^\dag \Psi)_{small}$ as function of $Z$ and $N$ for $Z\in [1,92]$
and $N\in [2,60]$. Only for very large $Z$ and very small $N$ this
contribution exceeds $1\%$.

\section{Wavepacket evolution}\label{evolution}

\subsection{Time scales}\label{tscal}

\begin{figure}[ht] 
\hspace{3mm}
 \resizebox{0.97\textwidth}{!}{\includegraphics{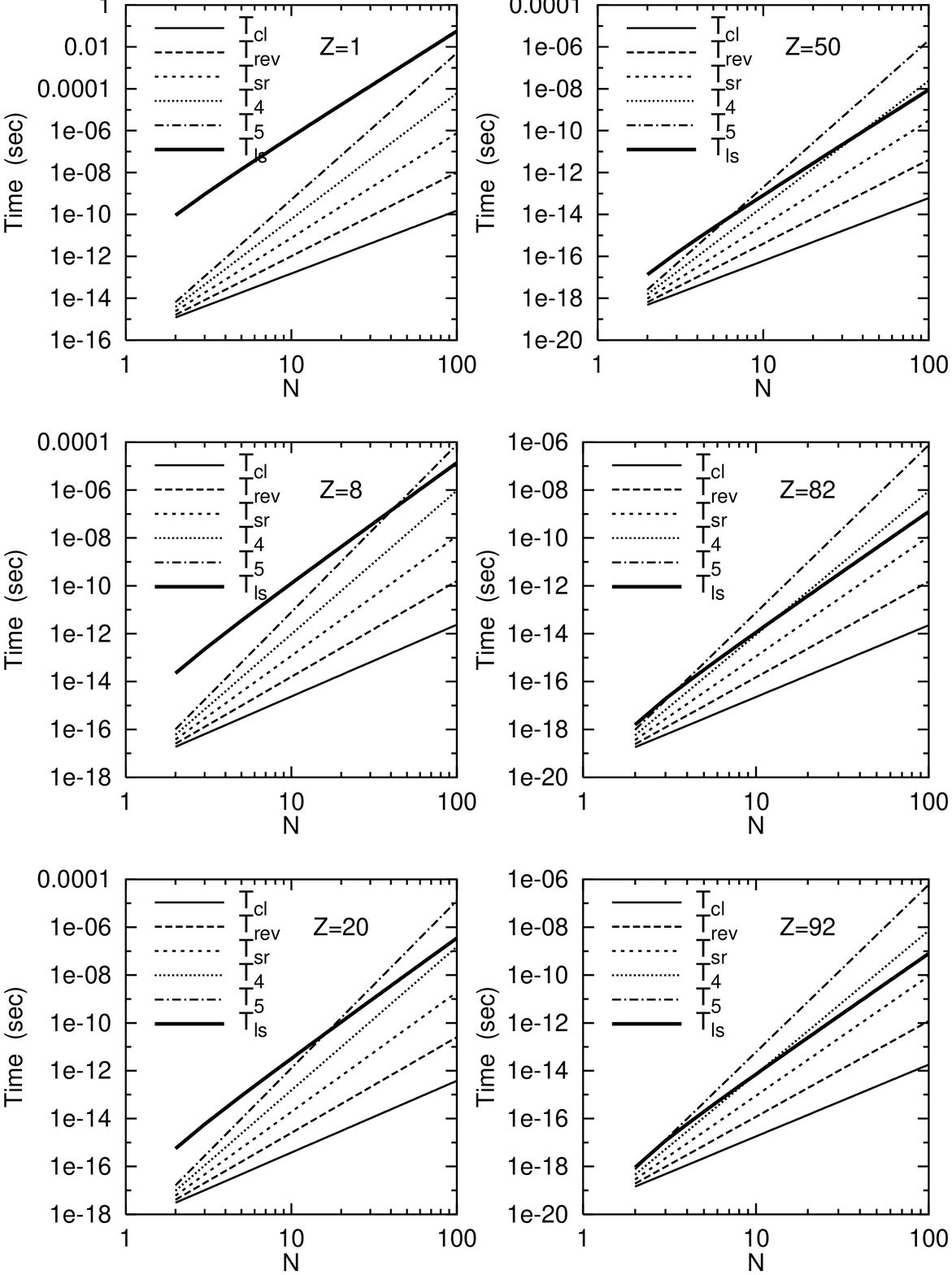} }
\caption{Characteristic times of the WP evolution as functions
(in log--log plot)
of the mean value of principal quantum number for different
values of nuclear charge (from $Z=1$ to 92). The period of spin--orbit
motion is given by the thick solid line.}
\label{times} 
\end{figure}

  We can define several characteristic times on the basis of the 
previous works about recurrent wave packets 
\cite{parker,averbukh,nauenberg,dacic,peres,bluhmsr,BKP}.
Writing the energy in function of
a single quantum number $n$ we define for $n=N$ a hierarchy of times
\begin{equation}\label{tk}
\frac{1}{k!} \left( \frac{\mbox{d}^k E}{\mbox{d}\,n^k}\right)_{n=N} 
= \frac{2\pi\hbar}{T(k)}
 \hspace{5ex} k=1,2,3,\ldots
\end{equation}
   For $k=1$ we obtain the classical Kepler time, for $k=2$ the revival
time, for $k=3$ the superrevival time etc\ldots
For hydrogenoid atoms those terms increase as $N^{k+2}$. 
These times can be compared to a characteristic spin
orbit time $T_{ls} = 2\pi/\omega_{ls}$ with $\omega_{ls}$ defined by
\begin{equation}\label{omgls}
          \hbar\omega_{ls} = E_{j_+=N-1/2}-E_{j_-=N-3/2} \;.     
\end{equation}
To first order we obtain 
\begin{equation}\label{omglsap}
           \omega_{ls} \approx \frac{(\alpha Z)^2}{N^5} \;.               
\end{equation}
Therefore the spin--orbit frequency  follows a variation parallel 
to the superrevival time with $k=3$.
Those times are plotted on Fig.~\ref{times} for $N\geq 2$
and for a large range of atoms (ions) including uranium.
For hydrogen $T_{ls}$ is orders of magnitude larger than all the 
other characteristic times $T(1)-T(5)$. 
If we want to observe spin--orbit 
effects in the picosecond regime we can concentrate on 
$N=20$ for $Z=92$. With such a choice  $T_{ls} \approx 1685\, T(1)$.
The other times are $T_2 =13.33\, T(1)$, $T_3 =200\, T(1)$,
$T_4 =3200\, T(1)$.
  The choice of $T_{ls}$ as a convenient time is justified by 
our plot of the autocorrelation function of Fig.~\ref{ac20} 
which shows an interesting oscillations up to 12 periods $T_{ls}$.

\subsection{Autocorrelation function}\label{autcor}

The autocorrelation function (often referred as the recurrence 
probability) contain a lot of information on wavepacket
evolution. It is a convenient measure of the degree of recurrences
and can be defined as:
\begin{equation} \label{acf}
 A(t) = \langle\Psi(0)|\Psi(t)\rangle =
\sum_{i=1}^{4}\, \langle c_i(0)|c_i(t)\rangle\, \:.
\end{equation} 
Using equations (\ref{ct1}-\ref{ct4}) one obtains
\numparts
\begin{eqnarray}\label{cc1}
\fl \langle c_1(0)|c_1(t)\rangle = \sum_l\, w_l^2 \left \{
 \left(a^2  + b^2 \frac{2l}{(2l+1)^2}\right) G_{j_+} 
 \rme^{-\rmi E_{j_+}t} \right. \\
 \hspace{8ex}+ \left. b^2 \frac{2l}{(2l+1)^2}
 \left[  G_{j_-} \rme^{-\rmi E_{j_-}t}
 - G_{j_{\pm}}(\rme^{-\rmi E_{j_+}t} + \rme^{-\rmi E_{j_-}t}) \right]  
 \right\}\;, \nonumber
\end{eqnarray} 
\begin{eqnarray}\label{cc2}
\fl \langle c_2(0)|c_2(t)\rangle = \sum_l\, w_l^2 \, b^2
 \left \{\frac{1}{(2l+1)^2} G_{j_+} \rme^{-\rmi E_{j_+}t} +
 \frac{(2l)^2}{(2l+1)^2} G_{j_-} \rme^{-\rmi E_{j_-}t} \right. \\
 \hspace{8ex} \left. + \frac{2l}{(2l+1)^2}
 G_{j_{\pm}}(\rme^{-\rmi E_{j_+}t} + \rme^{-\rmi E_{j_-}t})\right\} 
 \nonumber\;,
\end{eqnarray} 
\begin{eqnarray}\label{cc3}
\fl \langle c_3(0)|c_3(t)\rangle = \sum_l\, w_l^2 \left \{
 \left[ a^2 \frac{1}{2l+3}+b^2\frac{2}{(2l+1)(2l+3)}\right]
 F_{j_+} \rme^{-\rmi E_{j_+}t} \right. \\  
 \hspace{8ex} \left.
 + b^2\frac{2l}{2l+1}F_{j_-} \rme^{-\rmi E_{j_-}t} \right \} \nonumber\;,
\end{eqnarray} 
and 
\begin{eqnarray}\label{cc4}
\fl \langle c_4(0)|c_4(t)\rangle = \sum_l\, w_l^2 \left \{
 a^2 \frac{2l+2}{2l+3} + b^2 \frac{1}{2l+3}\right \}
 F_{j_+} \rme^{-\rmi E_{j_+}t} \;.
\end{eqnarray} 
\endnumparts
Terms $G_{j_+}, G_{j_-}, F_{j_+}, F_{j_-}$ and $G_{j_{\pm}}$
are radial integrals (for instance 
$G_{j_+}= \int_{0}^{\infty}r^2\,g_{j_+}^2 dr$ and
$G_{j_{\pm}}= \int_{0}^{\infty}r^2\,g_{j_+}g_{j_-} dr$ and so on) 
that can be easily expressed in terms of $\mit\Gamma$ functions.

\begin{figure}[htb] 
\hspace{10mm}
 \resizebox{0.9\textwidth}{!}{\includegraphics{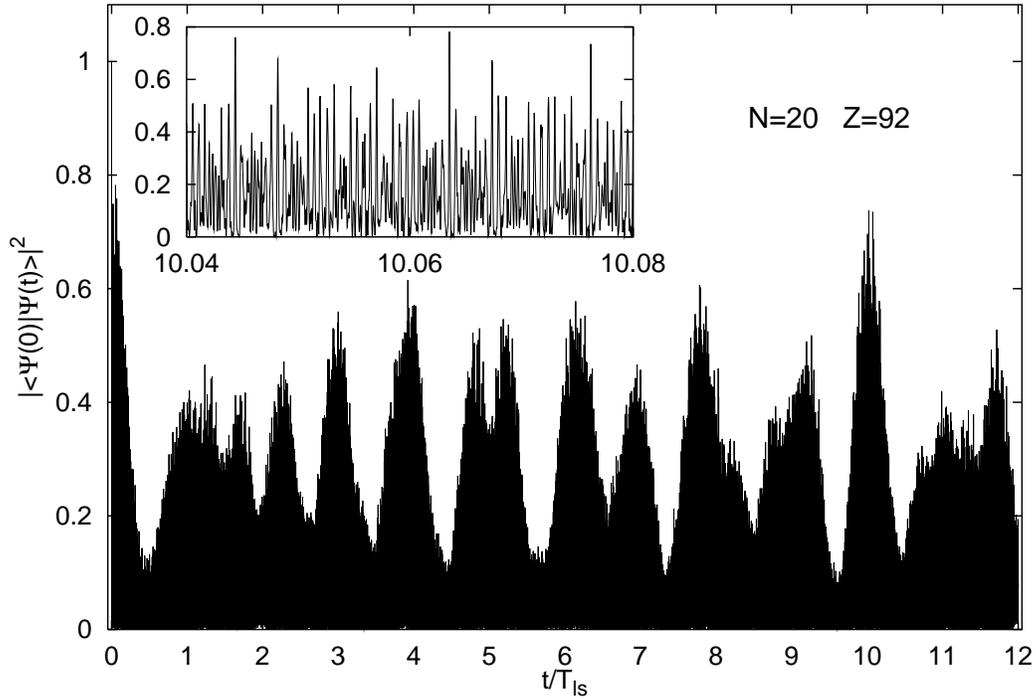} }
\caption{Square of the autocorrelation function for system with
$N=20$, $Z=92$ for long term evolution. In this case $T_{ls} 
\approx 1685\,T_{cl}$, where $T_{cl}=2\pi\,N^3/(Z\alpha)^2$ is
 a Kepler time for non-relativistic WP with the same $N$.
 Short term behaviour around the best revival is presented in
 the insert.}
\label{ac20} 
\end{figure} 
\begin{figure}[htb] 
\hspace{10mm}
 \resizebox{0.9\textwidth}{!}{\includegraphics{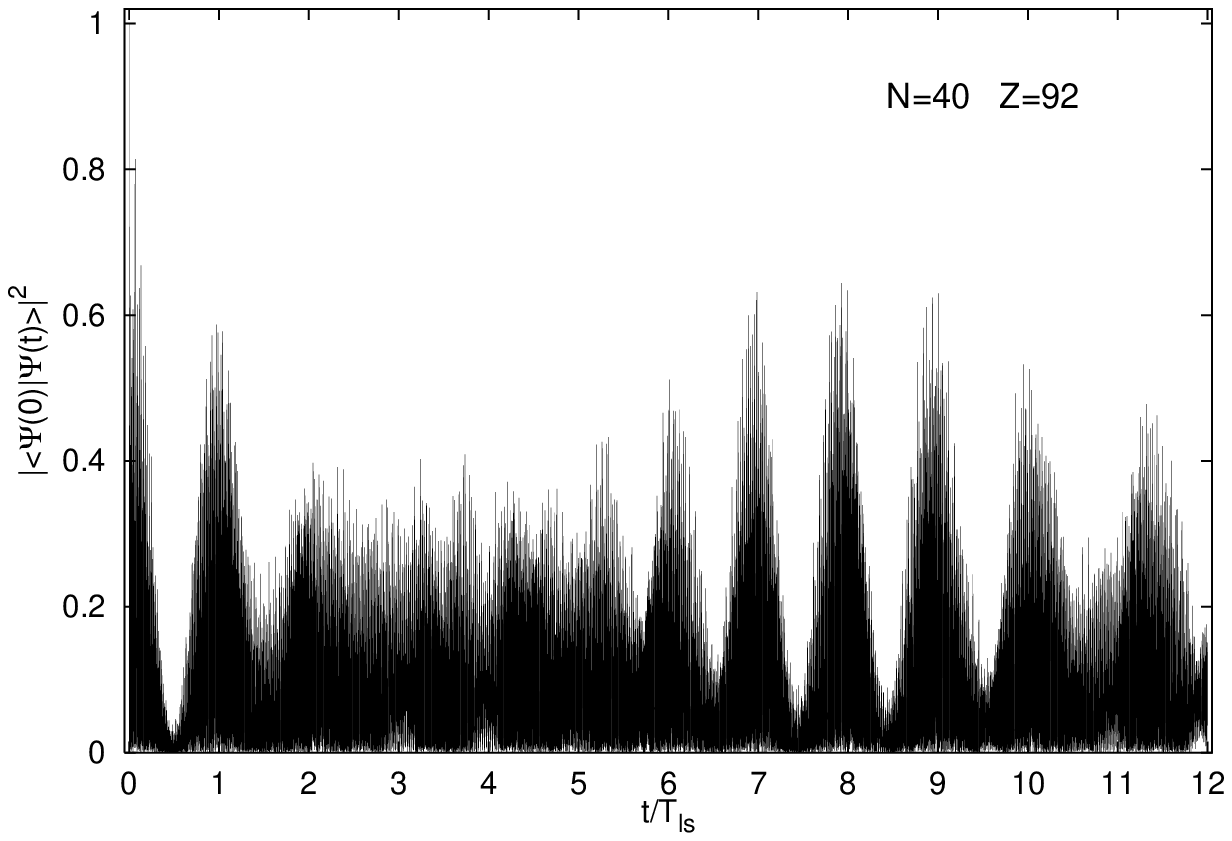} }
\caption{Square of the autocorrelation function for system with
$N=40$, $Z=92$ for long term evolution. In this case $T_{ls} 
\approx 6921\,T_{cl}$.}
\label{ac40} 
\end{figure} 

In the presentation of our results below we focus (as in our previous papers)
on the most interesting case, i.e. $a=b=1/\sqrt{2}$. This means that initial
direction of the spin is along the $Ox$ axis, in the orbit's plane. 
In Fig.~\ref{ac20} we present the square of the autocorrelation function
$|A(t)|^2$ as function of time for long term evolution ($T_{ls}$ units)
of the WP with $N=20$. Heavy ion system with $Z=92$ is considered with the full 
relativistic approach. Very regular revival structure connected with integer
values of time in $T_{ls}$ units is immediately recognized. This regularity
is somewhat distorted by a strong influence of $T_4$ periodicity (see right
bottom picture of Fig.~\ref{times}) but becomes even more evident for
larger $N$. Both the main part and the insert of Fig.~\ref{ac20} 
show very complicated shorter scale behaviour of the autocorrelation
function related to the three shorter time scales: $T_{cl}, T_{rev}, T_{sr}$,
known already from the non-relativistic studies \cite{bluhmsr}. 
The case with $N=40$ shown in Fig.~\ref{ac40} suggests that for
times $t \in [0,1.5]\,T_{ls}$ and $t \in [6,10]\,T_{ls}$ the $T_{ls}$ 
time scale is essential for the evolution of the WP.
It shows also that there exist a deep connection between the time
behaviour of the autocorrelation function and the average values of 
the spin operators as discussed in the next subsection.

\subsection{Expectation values of spin operators}\label{spinev}

Expectation values of spin operators are derived directly from
(\ref{ct1}-\ref{ct4}). One obtains:
\numparts 
\begin{equation}\label{sxt}
\fl \langle\sigma_x\rangle_t = \sum_l w_l^2 \left\{
 \frac{2ab}{2l+1}(G_{j_+}+2lG_{j_{\pm}}\,\cos \omega_l t) 
 -\frac{2ab}{2l+3}F_{j_+} \right \} +\delta\sigma_x \;,
\end{equation}
\begin{equation}\label{syt}
\fl \langle\sigma_y\rangle_t = \sum_l w_l^2 \left\{
 2ab\frac{2l}{2l+1}G_{j_{\pm}}\,\sin\omega_l t\right \} 
 +\delta\sigma_y \;,
\end{equation}
\begin{eqnarray}\label{szt}
\fl \langle\sigma_z\rangle_t = \sum_l w_l^2 \left\{
 a^2G_{j_+}+b^2\frac{2l-1}{(2l+1)^2}(G_{j_+}-2lG_{j_-})
 -b^2\frac{8l}{(2l+1)^2}G_{j_{\pm}}\,\cos \omega_l t \right.
 \nonumber\\ \left.
\fl  \hspace{15ex} +b^2\frac{2l}{2l+1}F_{j_-}
 -\left( a^2\frac{2l+1}{2l+3}+b^2\frac{2l-1}{(2l+1)(2l+3)}
  \right) F_{j_-} \right\} \;,
\end{eqnarray}
\endnumparts
 where $\omega_l=(E_{j_+}-E_{j_-})/\hbar$. Terms denoted by
 $\delta\sigma_x$ and $\delta\sigma_y$, which are only small
 contributions, stem from coupling 
 between states with $l$ or $j$ different by 2 units.
Their explicit expressions are:
\numparts 
\begin{equation}\label{dsx}
\delta\sigma_x = \sum_l K_l \cos\tilde{\omega}_l t \;,
\end{equation}
\begin{equation}\label{dsy}
\delta\sigma_y = \sum_l K_l \sin\tilde{\omega}_l t \;,
\end{equation}
\endnumparts
where 
\numparts 
\begin{equation}\label{hwtild}
 \hbar \tilde{\omega}_l = E_{j_+} - E_{(j+2)_-}  \;,
\end{equation}
\begin{equation}\label{Fprim}
 F' = \int_0^\infty r^2\, f_{j+} f_{(j+2)_-} dr \;,
\end{equation}
and
\begin{equation}\label{Kl}
 K_l = 2ab\, w_l w_{l+2}\, F'
 \sqrt{\frac{(2l+2)(2l+4)}{(2l+3)(2l+5)}} \;.
\end{equation}
\endnumparts

In a non-relativistic limit the terms containing the radial parts 
$f_{j_+}$ and $f_{j_-}$ cancel and the correction terms 
$\delta\sigma_x$  and $\delta\sigma_y$ tend to zero,
the same for $F_{j_+}$ and $F_{j_-}$.
For the harmonic oscillator with a constant
spin--orbit potential the radial integrals verify
\begin{equation}\label{Gjpm}
               G_{j_+}=G_{j_-}=1  \;.      
\end{equation}
The dynamics of the spin motion in such a case has been discussed by 
us in \cite{arv94} and \cite{arv95}. Using the dispersion law 
appropriate to a constant spin--orbit potential 
\begin{equation}\label{omgl}
                \omega_l=(2l+1)\omega_{ls}   
\end{equation}
we have obtained there the periodic oscillations between spin and orbital
angular momentum that we have called {\em spin--orbit pendulum}.

   In the present, more general case we see that both 
  $\langle \sigma_x \rangle$ and $\langle \sigma_z \rangle$
contain an additional constant shift with respect to the non-relativistic
expressions, coming clearly from the small components. 
In addition there are time dependent shifts due to the 
$\delta\sigma_x$  and $\delta\sigma_y$ terms. 
Numerically these shifts may be safely neglected. Therefore the most
interesting dependence comes from the use of the eigenvalues of the 
Dirac hamiltonian in the time behaviour that we will now discuss 
in more details. 

\begin{figure}[htb] 
\hspace{5mm}
 \resizebox{0.9\textwidth}{!}{\includegraphics{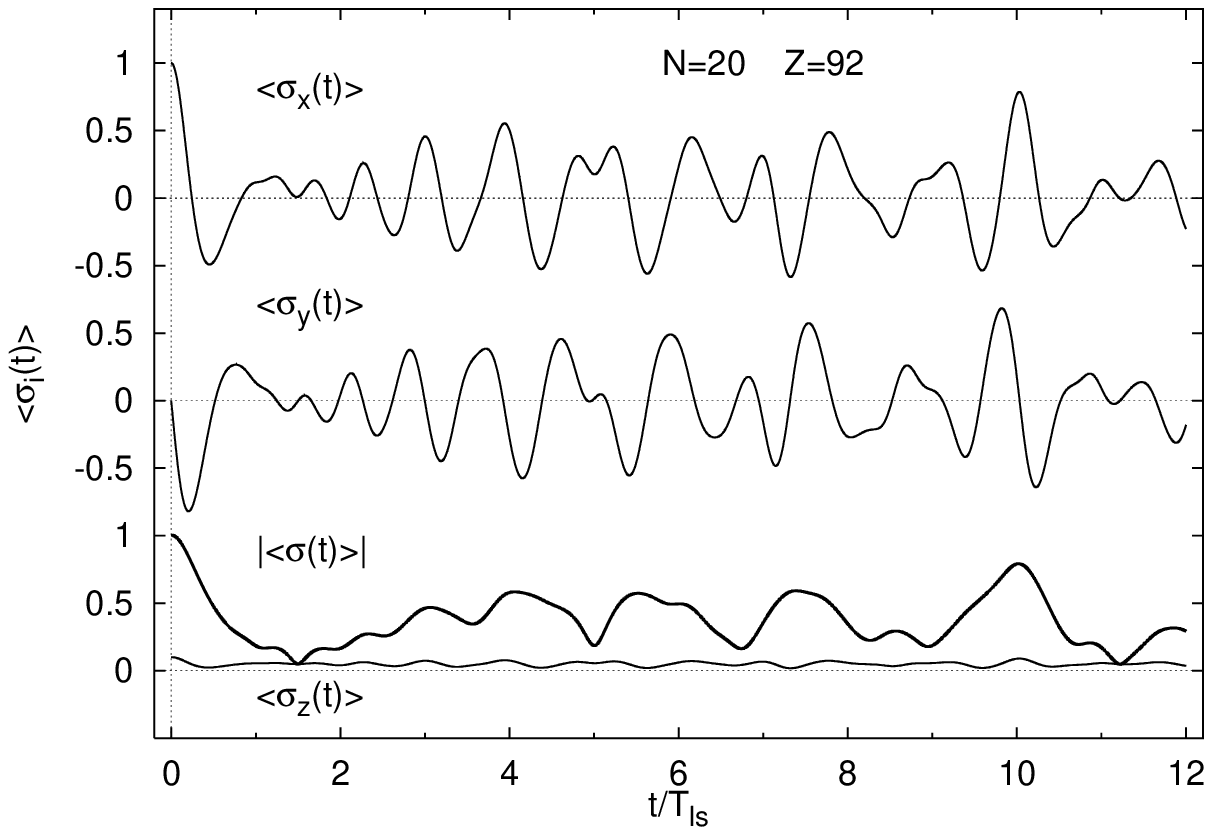} }
\caption{Time evolution of the average values of spin components 
$\langle \sigma_i \rangle$, $i=x,y,z$ 
for system with $N=20$, $Z=92$.}
\label{sa20} 
\end{figure} 
\begin{figure}[htb] 
\hspace{5mm}
 \resizebox{0.9\textwidth}{!}{\includegraphics{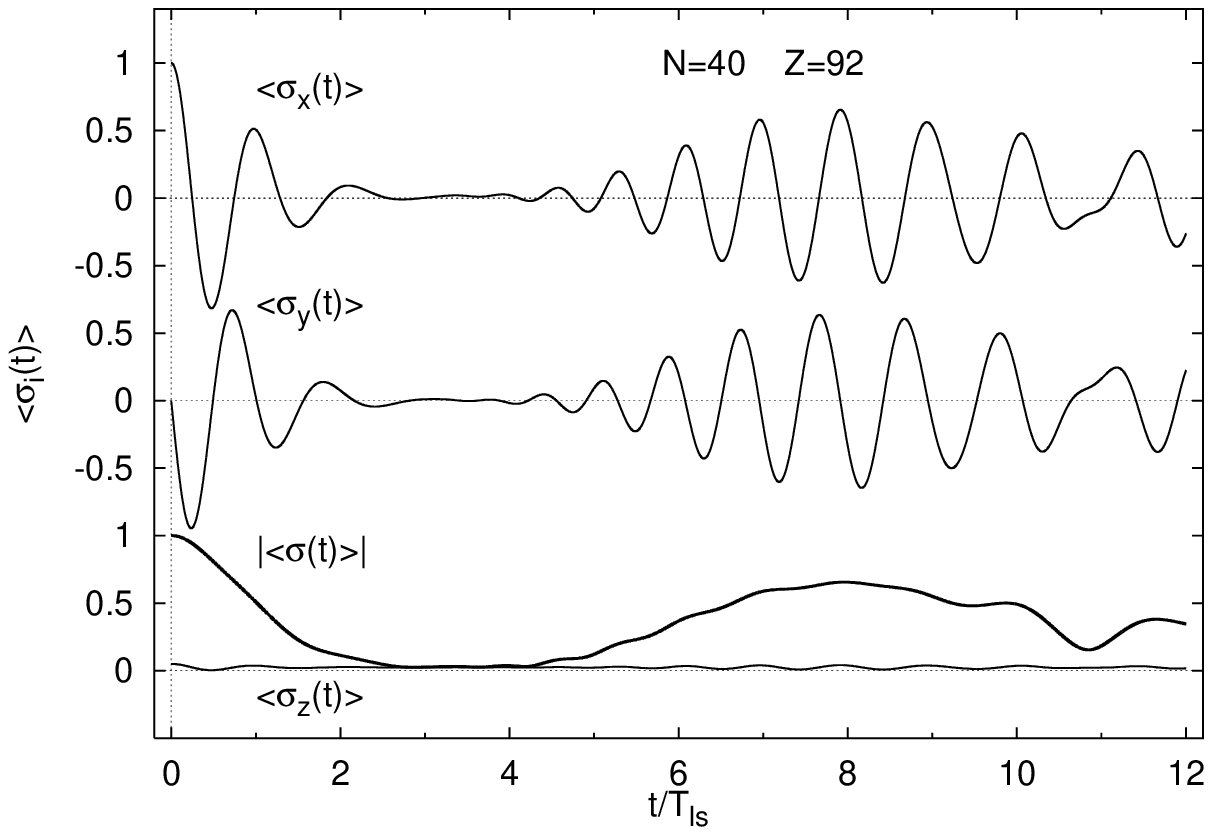} }
\caption{Time evolution of the average values of spin components 
$\langle \sigma_i \rangle$, $i=x,y,z$  
for system with $N=40$, $Z=92$.}
\label{sa40} 
\end{figure}

In Figs.~\ref{sa20} and \ref{sa40} we present the average values of the
spin operators and the length of the spin vector as functions of time
for cases $N=20$ and $N=40$ ($Z=92$). The latter case the most distinctly
reminds both non-relativistic {\em spin--orbit pendulum} for 
the HO+LS Hamiltonian \cite{arv94,arv95}
and the relativistic one for the DO \cite{roz99}. 
We clearly recognize time ranges for the spin collapse $(0\leq t\leq 2.5)$,
strong spin entanglement $(2.5\leq t\leq 4.5)$ and spin revivals
($6\leq t\leq 10$, all times in $T_{ls}$ units).
For the case $N=20$, shown in Fig.~\ref{sa20}, we see a similar behaviour
of the spin averages, distorted, however, much more due to stronger 
interferences with $T_4$ characteristic time. In cases discussed above
the ratio $T_4/T_{ls}$ is approximately equal to 1.9 for $N=20$ and
3.7 for  $N=40$.

\subsection{Spatial motion of circular WP}\label{spatial} 

\begin{figure}[tb] 
\hspace{15mm}
 \resizebox{0.84\textwidth}{!}{\includegraphics{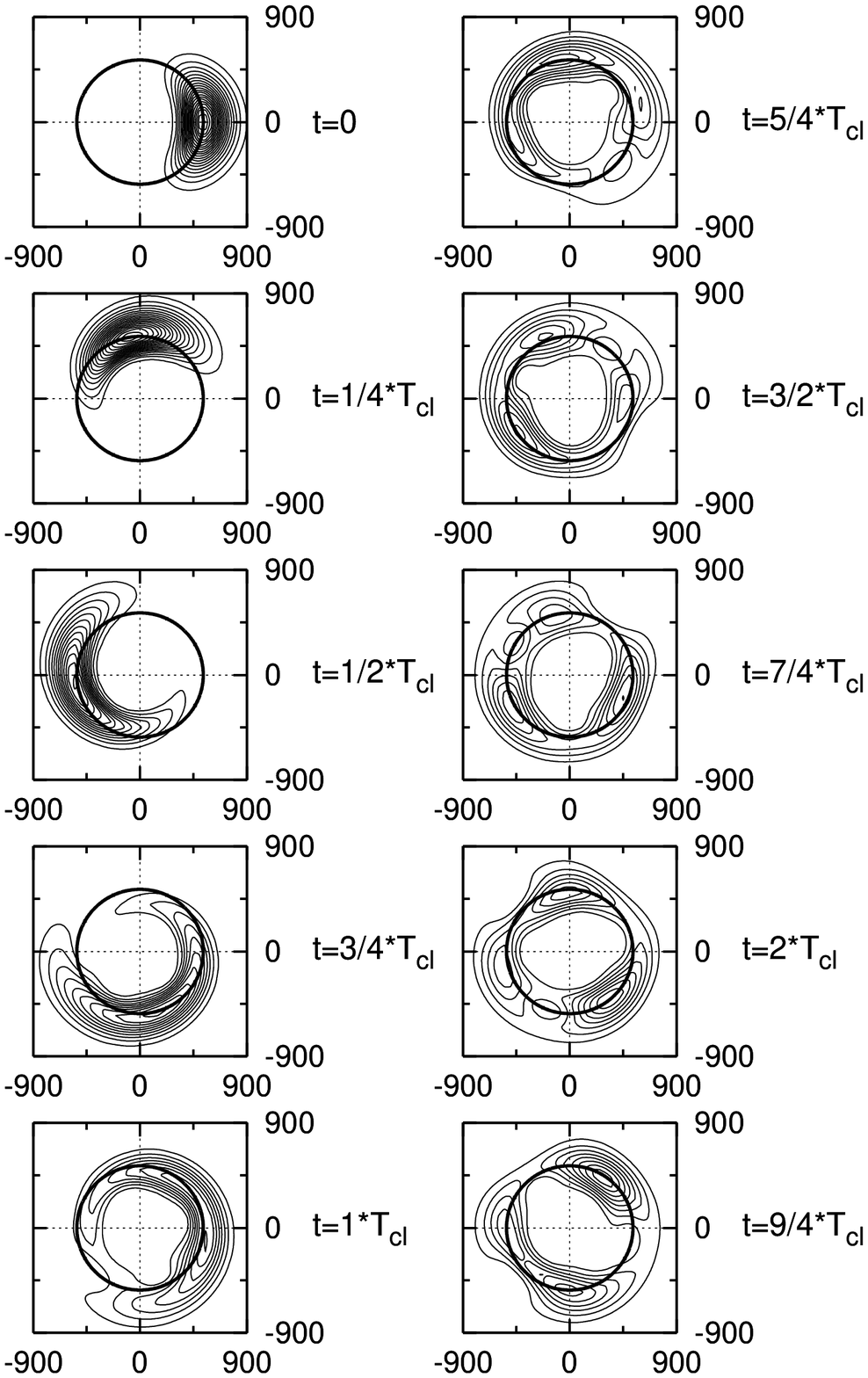} }
\caption{Spatial motion of the WP with $N=20$, $Z=92$ in the initial stage
of time evolution $t\in [0,9/4]T_{cl}$. The probability density 
$\Psi^\dag\Psi$ is plotted as function of the position on the plane of the
electron's classical orbit. The radius of the thick circle corresponds
to the initial distance of the WP's peak from the center of the potential.
Note that in this figure and all next plots the contour lines correspond to
the same values of the probability density. }
\label{xoy0} 
\end{figure}

Let us now discuss the motion of WP in space. A complex structure
of the autocorrelation function and the existence of several time scales
suggest that qualitatively similar spatial evolution can occur in different
time scales. Below we focus our presentation of the probability density
motion in two particular time scales: the shortest one, corresponding to
the Kepler period of classical electron motion on the circular orbit
with $n=N$, and the spin--orbit motion time scale.
The admixture of the components $|l,l-1\rangle$ in $|c_1(t)\rangle$
(\ref{ct1}) [as shown earlier the small components give negligible contributions to
$\Psi^\dag\Psi$] causes deviations of the WP motion from the equatorial plane,
known already from our previous non-relativistic studies \cite{roz96,roz96ac}.
The magnitude of these deviations decreases, however, when $N$ increases
and for $N\ge 10-12$ the maximum of the probability density moves very close
to the equatorial plane,  i.e. the plane of the classical orbit.
Therefore we limit our presentation of  $[\Psi^\dag\Psi ](t)$ to this plane
only. For the short time scale, $t\in [0,9/4]\,T_{cl}$,
it is shown in Fig.~\ref{xoy0}. The sequence of pictures exhibits the 
behaviour known from non-relativistic studies on WP in hydrogen atom
\cite{brown,nauenberg,BKP,parker,averbukh,dacic,peres}, namely circular
motion with spreading along the orbit and forming interference patterns 
that lead to fractional revivals.
n Fig.~\ref{xoy0} 
we display the full probability density
 $\Psi^\dag\Psi$, not the spin up and spin down components separately,
because for such short time scale no difference in their
motion is visible.

\begin{figure}[tb] 
\hspace{15mm}
 \resizebox{0.85\textwidth}{!}{\includegraphics{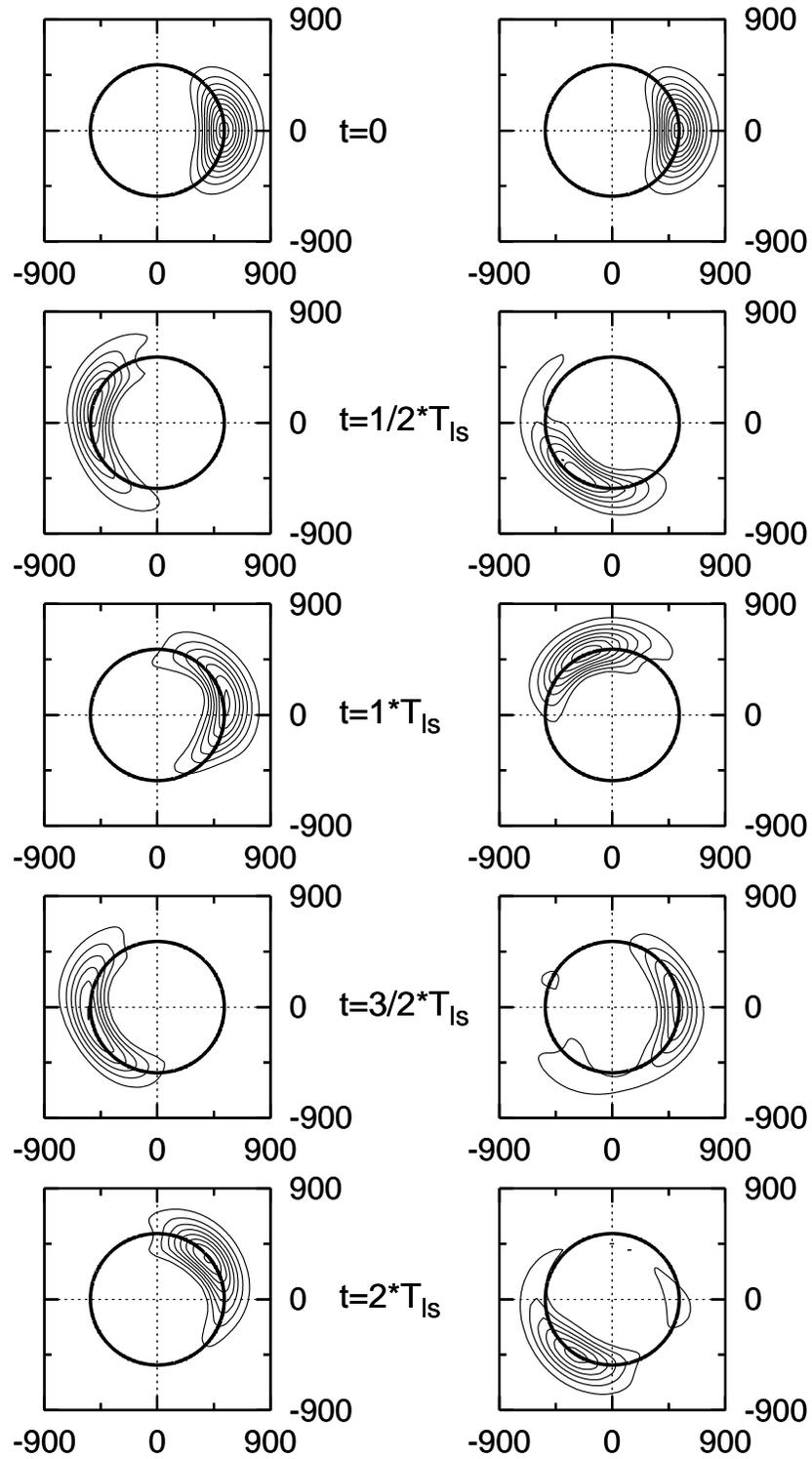} }
\caption{Shapes of the WP ($N=20, Z=92$) components with spin up 
(left column) and spin down (right column) during evolution in time scale 
measured in $T_{ls}$ units ($t\in [0,2]T_{ls}$). The different angular 
velocities of the motion of these sub-packets are clearly visible.}
\label{xoyT} 
\end{figure}

In order to see this difference one needs to observe WP in time scale
comparable to $T_{ls}$. In Fig.~\ref{xoyT} we present the sequence of shapes 
of the same WP ($N=20, Z=92$) for times $t=0,1/2,1,3/2$ and 2 (in $T_{ls}$
units). In these cases the presentation is divided into the two parts: 
in the left column spin up components of $\Psi^\dag\Psi$ are displayed,
whereas in the right column the spin down components are shown.
At the presented time instants, that are multiples of $T_{ls}/2$ both sub-packets
are quite well revived (localized) and move on a circle with different
velocities. 

\begin{figure}[htb] 
 \resizebox{0.99\textwidth}{!}{\includegraphics{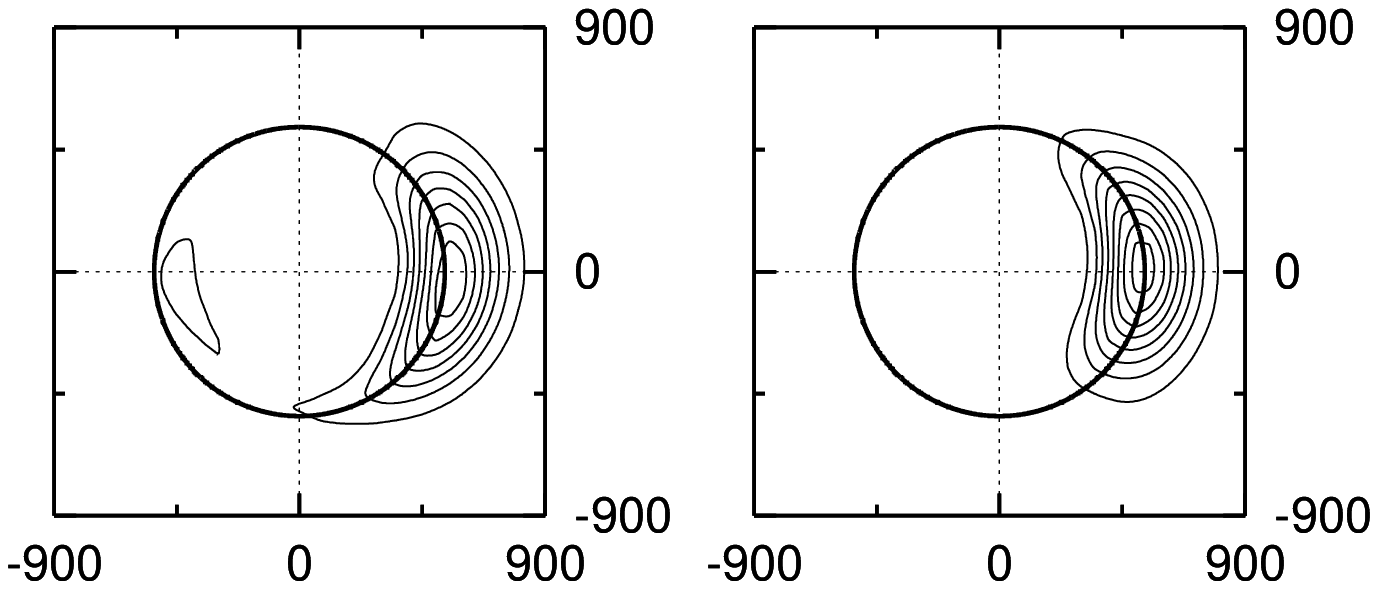} }
\caption{The best revival of the WP for the system with  $N=20$, $Z=92$ 
presented on the plane of the classical orbit. The time instant $t=10.063545\,
T_{ls}$ corresponds to the highest value of the autocorrelation function 
displayed in Fig.~\ref{ac20}. For comparison with the spin components
at $t=0$ see the upper row of the Fig.~\ref{xoy0}}
\label{xoyrev} 
\end{figure}

Fig.~\ref{xoyrev} 
shows (in the same convention) the shapes 
of WP's components with spin up (left) and spin down (right) for time
$t=10.063545\,T_{ls}$ ($N=20, Z=92$). As seen from Fig.~\ref{ac20}
this time corresponds to the highest value of the autocorrelation function
($|A(t)|^2\approx 0.8$) in the presented range of times. One sees from this
figure that the degree of the recurrences (of both sub-packets separately)
is indeed very high in this case.

\section{Conclusions}\label{conclusions}

We have discussed the full 3+1--dimensional evolution of the relativistic
wave packets in Coulomb potential. The main relativistic effect is the
appearance of the new time scale due to the spin--orbit coupling.
In general this effect has the same origin as {\em spin--orbit pendulum},
introduced by us both in non-relativistic \cite{arv94,arv95,roz96} 
and relativistic \cite{roz99} studies,
where harmonic oscillator potential with a strong spin--orbit coupling was 
considered. In the Coulomb potential (relativistic) case presented here
the time scale of the spin--orbit motion is large, therefore there is
practically no chance for observing it in such systems like hydrogen
or hydrogenoid atoms. However, this time scale decreases rapidly as function  
$1/Z^2$, when heavy ions with the single electron are considered as systems
for experimental studies. At $Z=82$, and even better $Z=92$ the $T_{ls}$
time becomes reasonably short and in principle experimental observations
of the spin--orbit effects in electron WP's motion may become possible
with existing techniques. This conclusion is based on the fact that the single
electron heavy ions with $Z=82,92$ have been already created and some of their
properties have been measured \cite{GSI}.
It is also possible to suitably adjust the two parameters $N,Z$ in
order to fit the energies of electron states and time scales of the motion
to ranges of action of available lasers that have to be used for creation 
of wave packets and analysis of their motion.

Comparing the phenomenon of the {\em spin--orbit pendulum} discussed in our
previous studies \cite{arv94,arv95,roz96,roz99} with the present Coulomb
potential case one sees the following differences. During their motion Coulomb 
wave packets experience several important time scales (only one 
is present in the HO case) that make the motion much more complex and
obscure the spin--orbit motion. Relativistic (nonlinear) dependence of the
eigenenergies on quantum numbers destroy exact periodicity of the motion.
  
\ackn
P.R. and M.T kindly acknowledge financial support (in part) of Polish Committee 
for Scientific Research (KBN) under the grant 2 P03B 143 14.

\end{document}